# NOMA in 5G Systems: Exciting Possibilities for Enhancing Spectral Efficiency

S. M. Riazul Islam, *Member, IEEE*, Ming Zeng, *Student Member, IEEE,* Octavia A. Dobre, *Senior Member, IEEE*

*Abstract*—This article provides an overview of power-domain non-orthogonal multiple access for 5G systems. The basic concepts and benefits are briefly presented, along with current solutions and standardization activities. In addition, limitations and research challenges are discussed.

## 1. Introduction

In recent years, non-orthogonal multiple access (NOMA) schemes have received significant attention for the fifth generation (5G) cellular networks [1]-[2]. The primary reason for adopting NOMA in 5G owes to its ability of serving multiple users using the same time and frequency resources. There exit two main NOMA techniques: power-domain and code-domain.[1] Power-domain NOMA attains multiplexing in power domain, whereas code-domain NOMA achieves multiplexing in code domain. This article focuses on power-domain NOMA, which henceforth is referred to as NOMA.

NOMA exploits superposition coding at the transmitter and successive interference cancellation

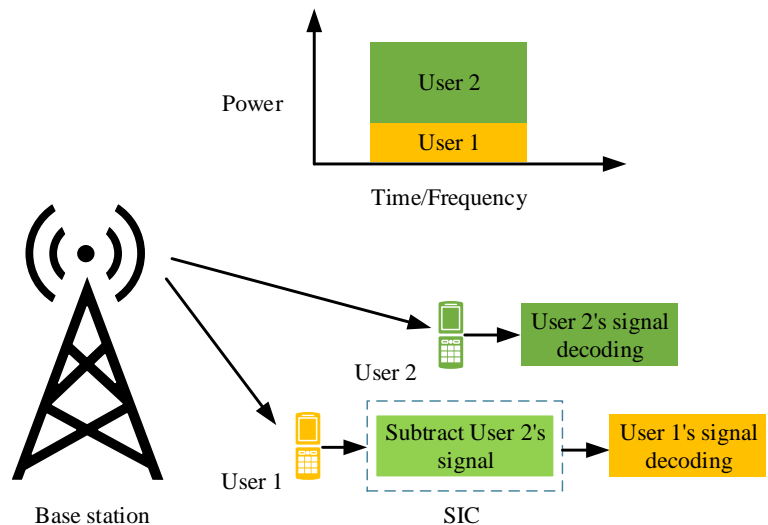

Fig. 1: Downlink NOMA in a single cell with one BS and two users.

---

[1] Examples of other types of NOMA are pattern division multiple access, multiuser shared access, and resource spread multiple access [1]-[2].

(SIC) at the receiver, thus multiplexing users in the power domain. As shown in Fig. 1, the base station (BS) sends the superposed signals to two users, where User 1 has higher channel gain than User 2. In NOMA, the user with higher channel gain and the user with lower channel gain are usually referred to as the strong user and the weak user, respectively. The strong user first subtracts the signal of the weak user through SIC, and then decodes its own signal; the weak user considers the signal of the strong user as noise and detects its own signal directly. With worse channel gain and more interference, the weak user is assigned more power in NOMA to ensure fairness.

## 2. Benefits of NOMA

NOMA dominates conventional orthogonal multiple access (OMA) in several aspects, such as: 1) It achieves superior spectral efficiency by serving multiple users at the same time and with the same frequency resource, and mitigating the interference through SIC; 2) It increases the number of simultaneously served users, and thus, it can support massive connectivity; 3) Due to the simultaneous transmission nature, a user does not need go through a scheduled time slot to transmit its information, and hence, it experiences lower latency; 4) NOMA can maintain user-fairness and diverse quality of service by flexible power control between the strong and weak users [3]; particularly, as more power is allocated to a weak user, NOMA offers higher cell-edge throughput and thus enhances the cell-edge user experience.

## 3. Existing NOMA Solutions

Saito et al. first uncovered the potential of NOMA for 5G cellular networks [4], and demonstrated that NOMA outperforms OMA in terms of capacity and user-fairness. Since then, researchers over the globe have started investigating how to transform the NOMA concept into the next generation radio access technique. Most early works on NOMA focused on single-input single-output (SISO), in which power allocation and user fairness are the main concerns. Power allocation in NOMA does not solely aim to maximize the sum rate, but considers sum rate and user fairness as a whole. This is because NOMA will assign all power to the strong user if the goal is to maximize the sum rate, and thus offers no gain over OMA. In [5], a dynamic power allocation scheme is proposed, which guarantees that the individual rates for both strong and weak users in NOMA are higher than the corresponding ones in OMA. In terms of user fairness, [6] considers the max-min data rate and min-max outage probability, respectively. Although the formulated problems are non-convex, low-complexity polynomial algorithms that yield the optimal solution are developed in both cases.

The performance of NOMA can be further enhanced by combining it with the multi-input multi-output (MIMO) technology. In MIMO-NOMA, users are paired into clusters, and NOMA is applied only among users in the same cluster. It is non-trivial to obtain the optimal user pairing, since an exhaustive search is needed. To relieve the computational burden, random pairing is adopted in [7]. In addition, based on the channel correlation and gain difference, a greedy user pairing algorithm is proposed in [8] to deliver near optimal performance. Once users are paired into clusters, a common precoding vector is shared by users in the same cluster, which transforms the MIMO channel into multiple parallel SISO channels. As a result, the superiority of NOMA over OMA still holds [9]. A general MIMO-NOMA framework is considered in [7], which eliminates the inter-cluster interference by adopting zero-forcing precoding and signal alignment-based detection. This work assumes perfect instantaneous channel state information (CSI), which may be impractical for MIMO scenarios. With statistical CSI knowledge, Sun et al. explore both optimal and low-complexity suboptimal power allocation schemes to maximize the ergodic capacity of a two user MIMO-NOMA system under the constraints of total transmit power and minimum rate of the weak user [10]. In [11], a NOMA scheme is proposed for a massive MIMO system by considering limited feedback channel.

Note that the works above are confined to a single-cell system. Recently, researchers have started to investigate the performance of NOMA in a multi-cell network, where inter-cell interference (ICI) is a major obstacle. To deal with the ICI in a two-cell MIMO-NOMA network, two coordinated beamforming approaches are proposed in [12]. In addition, NOMA is also promising for millimeter wave (mmWave) [13] and visible light communications [14].

## 4. Industry Trends and Standardization Status

There exists a burst of activity related to various non-orthogonal multiple access schemes for 5G cellular, which gets major players in the wireless industry excited. NOMA has been included in various White Papers on 5G, including those from ZTE Corporation, SK Telecom, and Mobile and wireless communications Enablers for the Twenty-twenty Information Society (METIS) consortium. As part of the 5G vision, DOCOMO and MediaTek are jointly researching to combine NOMA technologies from DOCOMO with multi-user interference cancellation technologies from MediaTek [15]. There exist noteworthy standardization activities on non-orthogonal multiple access schemes. Regarding NOMA, this was included in the study item on downlink multi-user superposition transmission for LTE by the 3rd Generation Partnership Project (3GPP) in its Release 13. NOMA was also considered in the initial study focusing on the 5G new radio (NR) of Release 14, which recognized that 5G NR should consider at least uplink non-orthogonal multiple access schemes, especially for massive machine type communications. Furthermore, the study of non-orthogonal multiple access schemes for 5G NR, including NOMA, will continue in Release 15 [16]. These studies involve the collaboration among different blue-chip vendors, such as DOCOMO, Huawei, Intel, Qualcomm, and Samsung.

## 5. Limitations of NOMA

Various limitations and implementation issues need to be addressed to exploit the full advantages of NOMA, such as: 1) Each user needs to decode the information of all other users with worse channel gains (which are in the same cluster) prior to decoding its own information [1], leading to additional receiver complexity and energy consumption compared with OMA; 2) When an error occurs in SIC at a user, the subsequent decoding of all other users' information will likely be carried out erroneously. This implies keeping the number of users in each cluster reasonably low to reduce the effect of error propagation; 3) To obtain the claimed benefits of power-domain multiplexing, a considerable channel gain difference between the strong and weak users is required. This intuitively restricts the effective number of user pairs, which in turn reduces the sum-rate gain of NOMA; 4) Each user needs to send back its channel gain information to the BS, and NOMA is inherently sensitive to the uncertainty in the measurement of this gain [3].

## 6. Research Challenges

Noticeably, a main research challenge is to overcome the limitations of NOMA discussed in the previous section. In addition, several other open issues need to be addressed.

• Firstly, as the density of BSs and consumer devices increases, the problem of ICI may become severe in multi-cell networks. The issue of interference mitigation might be even harder for heterogeneous networks. As such, determining techniques that combine useful interference cancellation and management approaches with NOMA is of noteworthy importance.

• Secondly, the integration of carrier aggregation (CA) with the NOMA-based system can further increase the data rates of the targeted users. However, what CA type is appropriate for NOMA solutions is not yet determined.

• Thirdly, developing low-complexity resource allocation algorithms is important, as they play a pivotal role for NOMA performance optimization. User pairing and power allocation, along with subband assignment are considered in [17].

• Fourthly, the integration of NOMA with other enabling technologies for 5G, i.e., massive MIMO and mmWave is of significance, and more efforts are required to explore NOMA solutions suitable for massive MIMO and mmWave networks; [18] presents an initial study in this direction.

• Finally, how to realize physical layer security for NOMA is an interesting research direction; [19] investigates this issue in large-scale networks.

## 7. Conclusion

This article illustrates that the superior spectral efficiency of NOMA is highly promising for 5G radio access. To date, NOMA has been investigated from various viewpoints, including resource allocation and fairness. The potential of NOMA is not limited to only SISO systems; its capacity can be further increased by applying NOMA in MIMO systems. The application of NOMA is also penetrable into other communication systems, including mmWave and visible light systems. To make NOMA more practical, its limitations such as error propagation and ICI in multi-cell networks should be overcome.

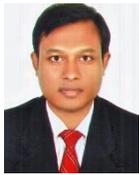

**S. M. Riazul Islam** (*Member, IEEE,* riaz@sejong.ac.kr) is an Assistant Professor with the Dept. of Computer Engineering at Sejong University, Korea. From 2014 to 2017, he worked at the Wireless Communications Research Centre, Inha University, Korea as a Postdoctoral Fellow. From 2005 to 2014, he was with the University of Dhaka, Bangladesh as an Assistant Professor and Lecturer at the Dept. of Electrical and Electronic Engineering. His research interests include signal processing for communications and enabling technologies for 5G and beyond.

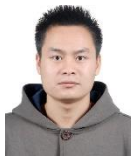

**Ming Zeng** (*Student Member, IEEE,* mzeng@mun.ca) received his B.E. and master's degrees from Beijing University of Post and Telecommunications, China. Currently, he is a Ph.D. student in the Faculty of Engineering at Memorial University, Canada. His research interests are in the area of wireless communications in general and resource allocation in emerging 5G systems in particular.

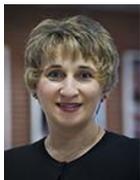

**Octavia A. Dobre** (*Senior Member, IEEE,* odobre@mun.ca) is a Professor and Research Chair at Memorial University, Canada. She was a Visiting Professor at Massachusetts Institute of Technology, and the recipient of a Royal Society Scholarship and Fulbright Fellowship. Her research interests include 5G enabling technologies, as well as optical and underwater communications among others. She published over 190 referred papers in these areas. Dr. Dobre serves as the Editor-in-Chief of the IEEE Communications Letters. She has been a senior editor and an editor with prestigious journals, as well as General Chair and Technical Co-Chair of flagship conferences in her area of expertise. She serves as Chair of the IEEE Communications Society WICE standing committee and is a member-at-large of the Administrative Committee of the IEEE Instrumentation and Measurement Society.